\documentclass[sigconf]{cidr-2026}
\usepackage{tikz}
\usetikzlibrary{arrows.meta,automata,positioning}
\usepackage{xcolor}
\usepackage{todonotes}
\AtBeginDocument{%
  }

\usepackage{float}

\usepackage{graphicx}
\usepackage{minted}
\usepackage{booktabs}
\usepackage{amsmath}
\usepackage{amsthm}

\theoremstyle{definition}

\newtheorem*{definition*}{Definition}

\usepackage{sparklines}
\usepackage[export]{adjustbox}
\usepackage{subfig}
\usepackage{xspace}
\usepackage{xcolor}
\usepackage{soul}
\usepackage{marginnote}
\usepackage{enumitem}
\usepackage[algoruled, linesnumbered, vlined]{algorithm2e}
\usepackage{setspace}
\usepackage{multirow}
\usepackage{multicol}
\usepackage{circledsteps}
\usepackage{tcolorbox}
\tcbuselibrary{skins, breakable}

\definecolor{oldpaperyellow}{RGB}{245, 236, 204}
\newcommand{\note}[1]{\noindent{\color{red}\textbf{#1}}}

\newcommand{\tinyskip}{\vspace{3pt}}
\newcommand{\mypar}[1]{\tinyskip\noindent\textbf{#1.}\xspace}

\newenvironment{myitemize}{%
\begin{itemize}[leftmargin=1em, itemsep=.1em, parsep=.1em, topsep=.1em,
    partopsep=.1em]}
{\end{itemize}}

\newenvironment{myenumerate}{%
\begin{enumerate}[leftmargin=1em, itemsep=.1em, parsep=.1em, topsep=.1em,
    partopsep=.1em]}
{\end{enumerate}}

\newenvironment{structure*}{\color{blue}\begin{myenumerate}}{\end{myenumerate}}

\sethlcolor{yellow}

\begin{document}

% Author-related
\newcommand{\ltf}[1]{\note{\color{olive}[\textsc{luthfi:} #1]}}

\title{The Pneuma Project: Reifying Information Needs as Relational Schemas to Automate Discovery, Guide Preparation, \\and Align Data with Intent}

\newcommand{\uchicago}{
  \institution{The University of Chicago}
  \city{Chicago}
  \country{USA}
}

\author{Muhammad Imam Luthfi Balaka}
\email{luthfibalaka@uchicago.edu}
\orcid{0009-0001-5324-7758}
\affiliation{\uchicago}

\author{Raul Castro Fernandez}
\email{raulcf@uchicago.edu}
\orcid{0000-0001-7675-6080}
\affiliation{\uchicago}

\begin{abstract}
Data discovery and preparation remain persistent bottlenecks in the data management lifecycle, especially when user intent is vague, evolving, or difficult to operationalize. The Pneuma Project introduces \textsc{Pneuma-Seeker}, a system that helps users articulate and fulfill information needs through iterative interaction with a language model–powered platform. The system reifies the user's evolving information need as a relational data model and incrementally converges toward a usable document aligned with that intent. To achieve this, the system combines three architectural ideas: context specialization to reduce LLM burden across subtasks, a conductor-style planner to assemble dynamic execution plans, and a convergence mechanism based on shared state. The system integrates recent advances in retrieval-augmented generation (RAG), agentic frameworks, and structured data preparation to support semi-automatic, language-guided workflows. We evaluate the system through LLM-based user simulations and show that it helps surface latent intent, guide discovery, and produce fit-for-purpose documents. It also acts as an emergent documentation layer, capturing institutional knowledge and supporting organizational memory.
\end{abstract}

\begin{CCSXML}
<ccs2012>
   <concept>
       <concept_id>10002951.10002952</concept_id>
       <concept_desc>Information systems~Data management systems</concept_desc>
       <concept_significance>500</concept_significance>
       </concept>
   <concept>
       <concept_id>10002951.10003317</concept_id>
       <concept_desc>Information systems~Information retrieval</concept_desc>
       <concept_significance>500</concept_significance>
       </concept>
   <concept>
       <concept_id>10003120</concept_id>
       <concept_desc>Human-centered computing</concept_desc>
       <concept_significance>500</concept_significance>
       </concept>
 </ccs2012>
\end{CCSXML}

\ccsdesc[500]{Information systems~Data management systems}
\ccsdesc[500]{Information systems~Information retrieval}
\ccsdesc[500]{Human-centered computing}

\keywords{Data Discovery, Data Preparation, Information Needs, Large Language Models, Human-Computer Interaction}

\maketitle

\section{Introduction}
\label{sec:introduction}

Large Language Models (LLMs) are poised to support the entire data management lifecycle, from collection to task execution~\cite{LLMDisruptDataManagement2023,DBGPT2024,zhang-etal-2025-tablellm,AutoDW2024}. Among the stages that have most resisted automation are data discovery, the task of identifying and retrieving documents relevant to a user's \emph{information need}~\cite{Aurum2018, FernandezDataDI2025}, and data preparation, the transformation of those documents into a usable form for downstream tasks. A key challenge lies in the nature of the user's information need: it is often vague, evolving, and difficult to express in a way that software can act on effectively. As LLMs become increasingly capable, the bottleneck shifts from executing tasks to helping users articulate their goals---their information needs---precisely enough to be fulfilled with available data.

Consider a practical example from our university's Finance department: \emph{``What impact will tariffs have on our organization?''} Answering this question requires addressing several steps:

\begin{myenumerate}
\item Discover relevant data sources, such as current tariff schedules and procurement records;
\item Define ``impact,'' which may include both direct (e.g., imported goods from tariffed countries) and indirect effects (e.g., tariffed components in otherwise unaffected imports);
\item Determine temporal scope: Is the user interested in projected impact next week, next fiscal year, or in a retrospective analysis?;
\item Integrate these data sources into a coherent document that supports meaningful interpretation.
\end{myenumerate}

To support such inquiries, a system must extract, make explicit, and operationalize the assumptions behind the question. This means surfacing the ``information need,'' which is the set of \emph{states of nature} required to answer the user's question~\cite{vod}. This articulated information need then guides discovery, integration, and preparation. In practice, these processes are rarely executed by a single user; modern organizations involve business analysts, data scientists, and data engineers collaborating in a semi-decentralized way to answer complex questions~\cite{DataScienceWorkerCollaborations2020}.

\textsc{Pneuma-Seeker} is designed to help users articulate and fulfill such information needs semi-automatically. A central insight behind \textsc{Pneuma-Seeker} is that an information need can be reified as a data model, specifically, a relational schema. The system then seeks to align that schema with available data, returning a document that satisfies the latent information need. Initially, neither the system nor the user may know exactly what this document looks like, but through interaction, both sides refine their understanding, e.g., the user may not ask to include the effect of direct and indirect tariffs initially, but expresses this explicitly after seeing an intermediate output. The user offers language-based feedback, and the system proposes increasingly aligned data models, converging over iterations toward one that meets the desired information need.

When the user asks \textsc{Pneuma-Seeker} the tariff question, the system examines a procurement database with dozens of tables, determines that tariff information is missing, retrieves relevant data from online sources, integrates the information into a tabular structure, and proposes a preliminary document for review. After two rounds of user feedback, the system converges on a schema and SQL query that estimates tariff exposure from German suppliers, matching the user's evolving understanding of ``impact.''

LLMs in \textsc{Pneuma-Seeker} serve as a bridge between language, which users employ to express intent, and data models, which the system uses to steer discovery and preparation. \textsc{Pneuma-Seeker} leverages recent advances in retrieval-augmented generation (RAG) \cite{LewisRAG2020} and agentic architectures to make this bridge actionable. To enable this functionality, \textsc{Pneuma-Seeker} introduces three generalizable contributions:

\begin{myitemize}
\item \textbf{Context Specialization.} LLMs struggle with large, diverse contexts in structured data tasks~\cite{lee2024learningreduceimprovingperformance,liu-etal-2024-rethinking,li-etal-2025-longtablebench}. \textsc{Pneuma-Seeker} addresses this by decomposing the workflow into subtasks with specialized contexts and narrow scopes.

\item \textbf{Conductor-Style Planning.} Rather than static, rule-based pipe-lines, \textsc{Pneuma-Seeker} employs a conductor component, an LLM-powered agent that assembles plans dynamically, based on real-time evidence of progress toward fulfilling the information need.

\item \textbf{Convergence Criteria.} \textsc{Pneuma-Seeker} treats the evolving data model as a shared state between user and system: the user refines it via language, and the conductor uses it to guide downstream modules. Convergence occurs when the document aligns with the user's latent information need.
\end{myitemize}

To evaluate \textsc{Pneuma-Seeker} without human subjects, we use \textsc{LLM Sim}, an LLM-based agent that acts as users, reacting only to system outputs. While not a full substitute for real studies, both quantitative and qualitative results show \textsc{Pneuma-Seeker} helps articulate and fulfill information needs.

\textsc{Pneuma-Seeker}'s design offers an important side effect: it acts as a mechanism for organizational knowledge capture. By prompting users to articulate their goals, it surfaces tacit information needs and latent dataflows. At scale, these interactions accumulate into emergent documentation, preserving tribal knowledge and institutional memory. As teams evolve, \textsc{Pneuma-Seeker} strengthens the resilience of internal data ecosystems by making this knowledge explicit and reusable.

\section{From Information Needs to Answers}

In this section, we define relevant concepts, survey the landscape of solutions that help users satisfy those needs, and present an interaction model that motivates the design of \textsc{Pneuma-Seeker}. We aim to highlight recurring challenges in how users articulate and refine their information needs, and to motivate a structured approach in which human information needs and system representations co-evolve toward the latent information need.

\subsection{Definitions}

\mypar{Information Need} An information need is the set of states of nature required to solve a data-driven task~\cite{vod}. This definition is general: an information need may include the features required for training a classifier, the schema to answer a SQL query, or the variables for causal inference. These states are often encoded in documents. For clarity, we assume one information need per task, though multiple valid representations may exist.

\mypar{Latent and Active Information Needs} The latent information need is the true set of states needed to solve a task, often initially unknown to the user. The active information need is the user's working hypothesis about what data is needed, which evolves through interaction and exploration to approximate the latent one.

\subsection{Landscape of Solutions}

\mypar{Document Retrieval} Once an active information need is specified, a wide array of technologies can help retrieve relevant documents. For example, web search engines map keyword queries to matching documents. In structured databases, SQL queries retrieve relations containing specific columns, assuming join paths exist. RAG systems enable LLMs to retrieve relevant text chunks from vector stores or databases before composing answers. These retrieval methods assume that the information need is reasonably well articulated, which is often the core bottleneck~\cite{Belkin:1980,kuhlthau1991inside,ExploratorySearch2009}.

\mypar{Identifying Information Needs} A separate but critical challenge is helping users formulate their information needs in the first place. Many user interfaces offer scaffolding to support this process: web autocomplete features leverage population-level priors; LLM interfaces often suggest next actions or clarifying prompts; enterprise data catalogs~\cite{JahnkeOttoEnterpriseDataCatalog2023} surface usage patterns (e.g., users querying table A and B often also use table C), aiding newcomers in forming expectations about what data is relevant.

\mypar{Representing Information Needs}
The notion of an information need has long been studied, especially in information retrieval (IR)~\cite{Belkin:1980,Saracevic1996Relevance}. In IR, an information need is often framed as the user's underlying goal, which a query only imperfectly expresses. In exploratory search, users engage in open-ended, investigative information seeking where goals evolve during the search process~\cite{Marchionini2006ExploratorySearch,WhiteRoth2009ExploratorySearch}. Similarly, in exploratory data analysis (EDA), the focus shifts to iterative refinement and hypothesis evolution, often guided by visualization~\cite{Tukey1977EDA,HeerShneiderman2012InteractiveDynamics}. In the Pneuma project we build on these lines of work to provide computational representations of information needs that human users and machines can co-evolve.

In this work, we reify an information need as a relational data model plus a SQL query over that model. This choice reflects our view that solving a data-driven task ultimately involves instantiating a structured document (a table or set of tables) that aligns with the user's intent. The system's job, then, is to identify the schema and query that materialize the latent information need.

\subsection{Aligning Data with Intent: A Model}

We propose the following model to guide system design. A human user seeks to solve a data-driven task. That task induces a latent information need, which, if represented as a document, would suffice to solve the task. However, the user may not initially know what that document looks like.

An abstract system, initially agnostic to the task, holds an internal state representing its evolving understanding of the information need. The interaction proceeds in iterations: the user communicates their active information need, and the system updates its state accordingly. After observing the system's output, the user revises their input, gradually steering the system closer to the latent information need. The interaction concludes when the system produces a document (or schema + query) that the user deems sufficient.

In contemporary practice, this system is not software alone—it is a composite of human roles and infrastructure: business analysts, data scientists, data engineers, domain experts, databases, interfaces, APIs, and more~\cite{DataScienceWorkerCollaborations2020}. With \textsc{Pneuma-Seeker}, we aim to unify these roles under a coherent software system that helps users identify, articulate, and materialize their information needs through iterative, language-guided interaction.

\section{The \textsc{Pneuma-Seeker} System}
\label{sec:solution Overview}

In this section, we describe \textsc{Pneuma-Seeker}, a system that helps users identify and fulfill their latent information needs. We first present its architecture and key design principles, followed by a detailed explanation of each component.

\subsection{Technical Contributions and Overview}
\label{subsec:architecture}
\textsc{Pneuma-Seeker} is designed around three technical insights:

\mypar{Context specialization} When working with structured data, an LLM is constrained by context size heterogeneity. The more heterogeneous the context, the more attention is spread across unrelated details. In addition, prompting an LLM with distinct \textit{roles} can help focus its behavior~\cite{LLMRolePlay2023}. Thus, \textsc{Pneuma-Seeker} adopts a multi-component architecture in which retrieval, integration, and orchestration are separated. Each component focuses on a single subtask, and the LLM receives only the information relevant to its assigned role.

\mypar{Dynamic planning with \textsc{Conductor}} The subtasks induced by context specialization must still be assembled into an end-to-end process of identifying and fulfilling a user’s information need. Rather than relying on a static, rule-based pipelines, \textsc{Pneuma-Seeker} uses \textsc{Conductor}, an LLM-powered agent that orchestrates the process based on real-time evidence of progress toward fulfilling the information need.

\mypar{Shared state for convergence} \textsc{Pneuma-Seeker} represents a user's active information need as a relational data model $(T,Q)$, where $T$ is a set of tables, and $Q$ is a sequence of SQL queries over $T$. The user and \textsc{Pneuma-Seeker} establish a feedback loop: the user describes their active information need as it changes to the system, the system updates $(T,Q)$, and the user reacts to those changes with additional feedback. The interaction ends when the user stops or when the active information need matches the latent information need (i.e., convergence).

\begin{figure}[h]
    \centering
    \includegraphics[width=1.0\linewidth]{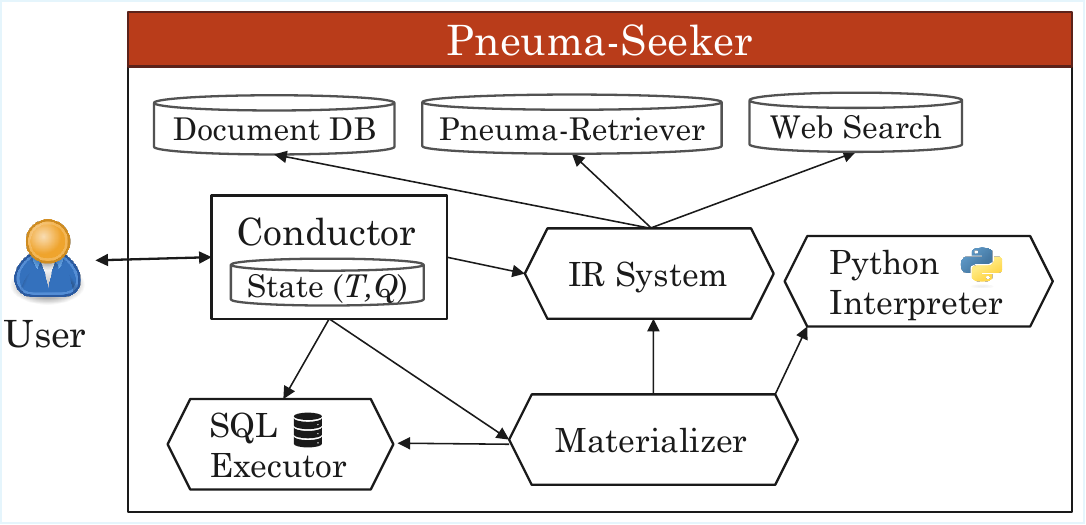}
    \caption{The Architecture of \textsc{Pneuma-Seeker}}
    \label{fig:architecture}
\end{figure}

These three insights directly shape \textsc{Pneuma-Seeker}'s modular architecture. Figure~\ref{fig:architecture} shows the architecture, which consists of multiple components: \textbf{(1) \textsc{Conductor}}, which orchestrates the process; \textbf{(2) \textsc{IR System}}, which retrieves relevant data; and \textbf{(3) \textsc{Materializer}}, which integrates and prepares data to form $T$. We describe them in detail below.

\subsection{\textsc{Conductor}}
\label{subsec:conductor}
\textsc{Conductor} drives \textsc{Pneuma-Seeker} towards convergence by selecting actions on the fly to align $(T,Q)$ with a user's active information need. At the moment, it selects actions one at a time, but there has been some research on parallelized LLM planning (e.g., \cite{zhang2025planovergraphparallelablellmagent,zhu-etal-2025-divide}). In \textsc{Conductor}, an action is any of the following:

\begin{myitemize}

    \item \textbf{Internal reasoning.} \textsc{Conductor} engages in internal reasoning (inspired by ReAct~\cite{yao2023react}), where we prompt the LLM so that it is able to evaluate the current state $(T,Q)$, retrieved data from \textsc{IR System}, and a user's most recent feedback, to decide the best next action(s). For example, suppose the user clarifies that determining the effect of a new tariff requires knowing the previous active tariff. \textsc{Conductor} might reason: \textit{``My current query computes price change using only the new tariff percentage. I should retrieve the previous tariff percentage and update the final percentage to $(\text{new\_tariff}-\text{previous\_tariff})$}.''
    
    \item \textbf{Tool call.} \textsc{Conductor} can directly call tools, including \textsc{IR System}, \textsc{Materializer}, and \textsc{SQL Executor}, or invoke them after first identifying the need through internal reasoning. For example, if it determines that the previous active tariff is required, it may issue a retrieval request to \textsc{IR System}: \textit{“Retrieve the previously active tariff for the region.”} Similarly, if it decides that $T$ should be materialized, it may invoke \textsc{Materializer} with a note such as: \textit{``Materialize $T$, a table containing relevant procurement data with new and previously active tariff information for the supplier country.''}
    
    \item \textbf{State modification.} \textsc{Conductor} can update the state $(T,Q)$ to revise interpretations of the user’s active information need. It may modify only $T$, only $Q$, or both. For example, after determining that $Q$ should account for the previous active tariff, it may update $Q$ to look like this: \textit{[``SELECT price * (1 + new\_tariff - previous\_tariff) AS new\_price FROM procurement\_data'']}.
    
    \item \textbf{User-facing communication.} \textsc{Conductor} can interact directly with the user to explain actions taken, summarize the current state, ask clarifying questions, or propose next steps. For instance, after executing the queries in $Q$, it may inform the user: \textit{``I have executed the queries in $Q$. The new price for items bought from supplier 12345 in Germany is 5\% higher, with an average increase of \$5,125.''}
\end{myitemize}

By design, \textsc{Conductor} operates without fixed procedural rules, aside from essential dependencies (e.g., $T$ must be materialized before executing $Q$). \textsc{Conductor} limits the number of consecutive actions to a fixed value $i$, which we set to $i=5$ based on empirical observation that the LLM rarely hits this limit during our evaluation. This limit is intended to prevent $(T,Q)$ from moving away from the latent information need before user feedback can correct it, while also avoiding long autonomous runs that keep users waiting. In addition, we instruct \textsc{Conductor} to end each sequence of actions with a user-facing message whenever possible. If the action limit is reached without producing a user-facing message, the system interrupts and forces \textsc{Conductor} to do so.

When interacting with the user, \textsc{Conductor} grounds its decisions on data retrieved from \textsc{IR System}, rather than relying solely on assumptions (e.g., assuming we have suppliers with ID 12345 in a procurement table). This avoids producing unrealistic $(T,Q)$ states that cannot be materialized. For instance, if the user requests an analysis of the impact of a new tariff for a specific supplier in 2019, but \textsc{IR System} only retrieves tariff records from 2020 onward, \textsc{Conductor} can detect the gap and either search alternative sources or notify the user, rather than spending multiple actions building and materializing a $(T,Q)$. This prevents wasted effort and keeps the interaction focused on attainable outcomes.

\begin{figure*}[t]
    \centering
    \includegraphics[width=0.95\linewidth]{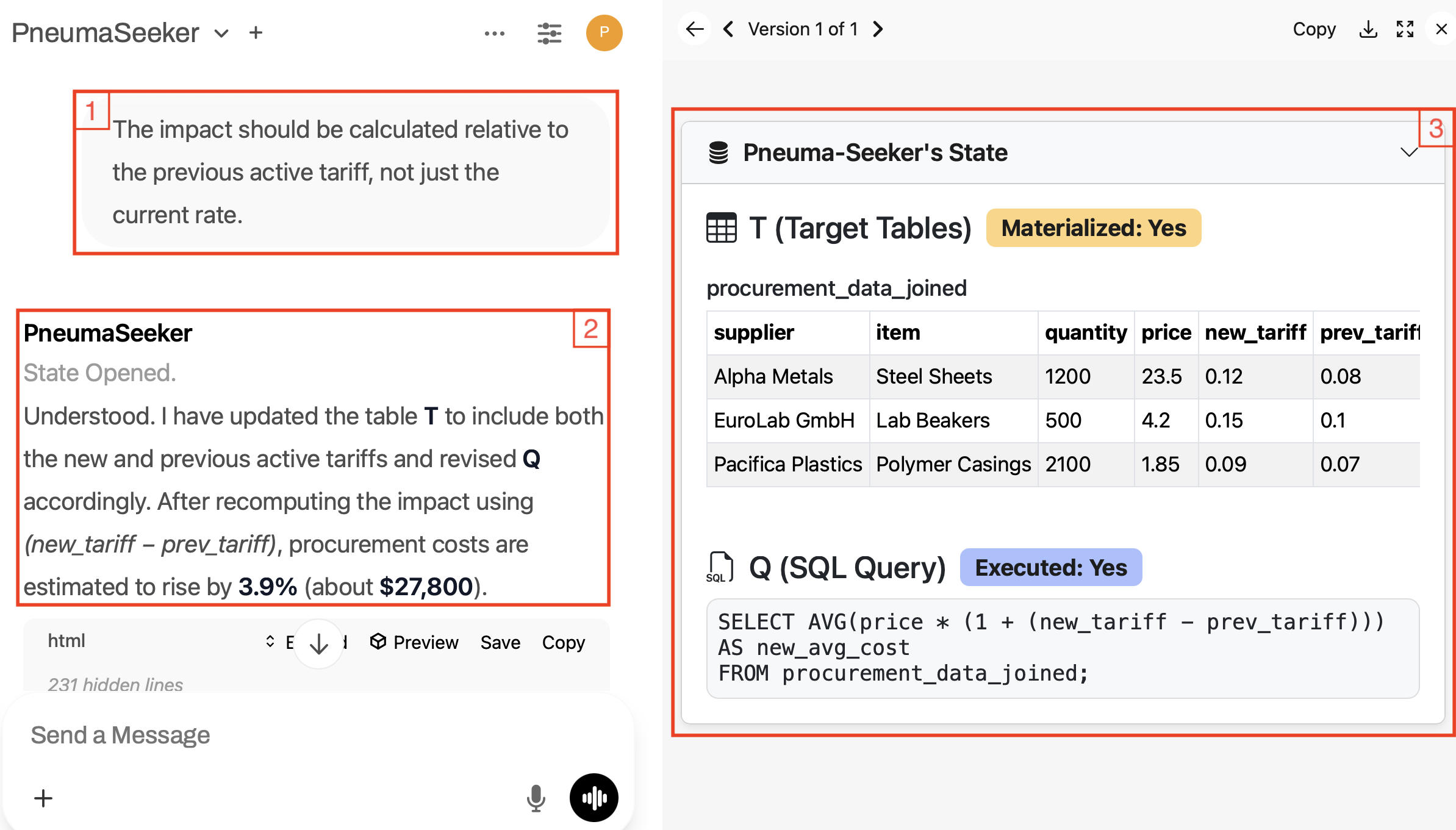}
    \caption{Interface of \textsc{Pneuma-Seeker}, showing: [1] User Query (Clarification), [2] User-Facing Message, and [3] State View Page $(T,Q)$. Note: the numbers and values of $T$ shown here are not real for privacy reasons.}
    \label{fig:state_view}
\end{figure*}

% \notera{mentions figure 2, but figure 2 is two pages in the future. bring figure 2 closer to this first mention}
Throughout an interaction, \textsc{Pneuma-Seeker} surfaces not only user-facing responses but also the current state $(T,Q)$ to the user. We display the interface in Figure~\ref{fig:state_view} (as of August 2025). Displaying the state (box~3; sample rows are shown for $T$) alongside the chat interface (box~1 and box~2) allows users to spot subtle mismatches between their intent and the system's evolving model and correct them. Continuing the tariff-impact example, suppose $T$ contains procurement data but lacks the \verb|country| attribute entirely, even though the user only cares about German suppliers. Without this attribute, $Q$ cannot filter the data correctly, and the final computation would be misaligned with the user's intent.

\subsection{\textsc{Information Retrieval (IR) System}}
\label{subec:ir_sys}
\textsc{IR System} supports \textsc{Conductor} and \textsc{Materializer} by retrieving relevant data from multiple sources. It abstracts heterogeneous retrieval format, such as tables and text, into document objects. This uniform representation allows new retrievers to be added without changing the rest of \textsc{IR System}'s design. Nevertheless, both \textsc{Conductor} and \textsc{Materializer} knows about the kind of data available in the system: currently tables, domain knowledge, and web pages, which are handled by the following three retrievers in our current implementation:

\begin{myitemize}
\item \textbf{\textsc{Pneuma-Retriever}~\cite{BalakaPneuma2025}}, a state-of-the-art table discovery system with a hybrid index, combining an HNSW~\cite{HNSW2018}-based vector store and a BM25~\cite{RobsertsonBM252009}-based inverted index for efficient table search.
\item \textbf{Document Database}, which uses \textsc{Pneuma-Retriever}'s indexer to store domain knowledge. This enables cross-user knowledge transfer. For example, if one user specifies that estimating tariff impacts requires accounting for both direct and indirect tariffs, subsequent tariff-related queries can leverage that insight. \textsc{Pneuma-Seeker} automatically captures knowledge from user interactions and save it to Document Database, inspired by prior work (e.g.,~\cite{park-etal-2021-scalable,CaptureKnowledgeUrban2025,liu-etal-2025-user}).
\item \textbf{Web Search}, which provides a thin interface to external search engines for general or up-to-date information lookup.
\end{myitemize}

With the current implementation, \textsc{Pneuma‑Seeker} can already handle complex users' information needs that combine structured and unstructured sources. Building on this design, we will extend \textsc{IR System} to accept direct feedback on retrieved documents from \textsc{Conductor} or \textsc{Materializer}.

\subsection{Materializer}
\label{subsec:materializer}
\textsc{Materializer}'s sole purpose is to populate $T$ with data, possibly involving integration of multi-source data from \textsc{IR System}. \textsc{Materializer} is separate from \textsc{Conductor}, reflecting the \textbf{context specialization} principle: \textsc{Materializer} operates only on context relevant to data integration and transformation, without being distracted by orchestration details. \textsc{Materializer} also considers $Q$, so it understands filters in the queries. For example, if a query in $Q$ expects a \verb|date| column to be of format ``yyyy-mm-dd,'' while the column actually lists its values with format ``Month Day, Year,'' \textsc{Materializer} will transform the values of this column by producing Python code.

Materializer’s toolkit includes a DuckDB~\cite{RaasveldtDuckDB2019}-based \textsc{SQL Executor} for standard relational operations (joins, unions, etc.) and a Python interpreter equipped with Pandas~\cite{mckinney-proc-scipy-2010} and NumPy~\cite{HarrisNumpy2020}. Similar to \textsc{IR System}, \textsc{Materializer} is designed to be extensible, allowing new operators to be incorporated easily, including semantic relational operations (e.g., as defined by LOTUS~\cite{PatelSemanticOperator2024}).

Through its prompts, \textsc{Materializer} is aware of all retrieved documents from \textsc{IR System} and can leverage them to form SQL queries or Python code. For example, if \textsc{Materializer} retrieves two documents representing (1) the latest tariffs from Web Search and (2) internal procurement tables from \textsc{Pneuma-Retriever}, then it can generate Python code that combines both documents (e.g., tariff information becomes a new column in the table). Errors may occur (e.g., floating-point operations on columns with null values), so the respective tool analyzes these errors and provides feedback to \textsc{Materializer} to fix the generated queries or code.

We plan to explore interaction models in which \textsc{Conductor} provides incremental feedback to \textsc{Materializer} after every $n$ operations, mirroring the way users interact with \textsc{Conductor}. This additional supervision could help prevent \textsc{Materializer} from drifting away from the intended $T$, especially in long integration pipelines.

\subsection{Dynamic vs. Static Pipelines}
In this section, we reflect on previous designs we experimented with and justify \textsc{Pneuma-Seeker}'s design based on that experience.

We design both \textsc{Conductor} and \textsc{Materializer} is significantly more adaptable for incorporating more actions/tools (and hence cover more use cases) compared to the static pipeline we initially experimented with---define $(T,Q)$, retrieve top-$k$ tables, filter and integrate the tables via relational operations, and prune the integration results down to those defined in $T$. While such a static pipeline can cover a variety of use cases, it becomes harder to update as we incorporate more actions/tools. Whenever a step in the original pipeline is insufficient (e.g., lacking the actions needed to cover a new use case), the entire pipeline must be revisited.

For example, in our tariff-impact scenario, there may be cases where the correct analysis requires joining procurement data with tariff records, but the sources do not share a common identifier (e.g., \texttt{supplier\_id}). In such cases, the system needs to employ more flexible relational operations (e.g., semantic or fuzzy joins) that were not part of the original pipeline. Incorporating such an operation into a static, predefined pipeline may require significant engineering effort, since its position in the pipeline must be chosen carefully and its assumptions must align with both upstream and downstream operations.

In other words, generalizing a static pipeline to support a wide variety of use cases is possible, but every extension increasingly requires substantial effort. As more actions/tools are added, these integration costs compound. In contrast, with \textsc{Conductor} or \textsc{Materializer}, we simply define the new operation or tool, and it naturally fits into their action spaces.

\subsection{\textsc{Pneuma-Seeker} in Action}
To illustrate how the components work together, we revisit the finance department's broad question from Section~\ref{sec:introduction}: \textit{``What impact will tariffs have on our organization?''}

\noindent During the interaction, the user adds a key clarification: \textit{``Impact should be calculated relative to the previous active tariff, not just the current rate.''}

\textsc{Conductor} first engages in \textbf{internal reasoning}, recognizing that it needs both the new and previous active tariff information to compute the impact correctly. It then issues a tool call to \textbf{\textsc{IR System}}, which returns procurement tables from \textsc{Pneuma-Retriever} and tariff information from the Web Search interface. Based on this, \textsc{Conductor} defines $T$ as a single table combining procurement data with tariff information. It summarizes the retrieved sources and asks the user: \textit{``I found procurement records and historical tariff information for the relevant suppliers. I've designed $T$ to combine the tariff information with the procurement data. Should I materialize $T$ now?''}

Once the user confirms, \textsc{Conductor} calls \textbf{\textsc{Materializer}} to join the procurement and tariff data, thereby populating $T$. With $T$ materialized, \textsc{Conductor} performs a \textbf{state modification} to update $Q$ so that it computes cost impact relative to the previous tariffs:

\begin{verbatim}
["SELECT AVG(price * (1 + (new_tariff - prev_tariff))) 
 AS new_avg_cost FROM procurement_data_joined"]
\end{verbatim}

\textbf{\textsc{SQL Executor}} runs the updated query and returns an estimated x\% increase in procurement costs (about \$y) across all goods.\footnote{The actual numbers are not disclosed for privacy reasons.} \textsc{Conductor} reports this information to the user through \textbf{user-facing communication}:

\textit{``If the new tariffs take effect, procurement costs are estimated to rise by x\% (about \$y), calculated relative to the previous rates.''}

At this point, the user may further narrow the scope, e.g., to only consider lab equipment and suppliers from Germany, trigger another refinement cycle.

\section{Evaluation}
\label{sec:eval}

In this section, we offer preliminary evidence of \textsc{Pneuma-Seeker}'s ability to help users articulate information needs and solve data tasks by answering the following questions:

\begin{myitemize}
    \item \textbf{RQ1}: Can the user reach their underlying information need by interacting with \textsc{Pneuma-Seeker}?
    \item \textbf{RQ2}: Given a specific information need, can \textsc{Pneuma-Seeker} address it accurately?
\end{myitemize}

\noindent{\textbf{Evaluation Workload.}} We evaluate \textsc{Pneuma-Seeker} on the archaeology and environment datasets from KramaBench~\cite{LaiKramaBench2025}, with Web Search disabled to prevent leaking benchmark information from the internet. These datasets are associated with 12 and 20 questions, respectively. Table~\ref{tab:stats} shows the datasets' characteristics.

\begin{table}[ht]
\centering
\caption{Characteristics of the Datasets}
\label{tab:stats}
\begin{tabular}{lcccc}
\toprule
\textbf{Dataset} & \textbf{\# Tables} & \textbf{Avg. \#Rows} & \textbf{Avg. \#Cols} \\
\midrule
Archeology & 5 & 11,289 & 16 \\
Environment & 36 & 9,199 & 10 \\
\bottomrule
\end{tabular}
\end{table}

Each benchmark question is a latent information need. We use an LLM (\verb|GPT-4o|) to simulate a domain expert (\textsc{LLM Sim}) interacting with the system. Starting from a broad prompt, \textsc{LLM Sim} iteratively refines its active information need based on the system's outputs. For example, given the latent question:

\textit{``What is the average Potassium in ppm from the first and last time the study recorded people in the Maltese area? Assume that Potassium is linearly interpolated between samples. Round your answer to 4 decimal places.''}

\noindent The initial query is:

\textit{“I’m curious to dive into the historical data from the Maltese region. Could you help me get an overview of the different variables we have for past studies?”}

\noindent Importantly, convergence is not guaranteed: \textsc{LLM Sim}'s active information need may never fully match the latent one. For reference, we display the prompt provided for \textsc{LLM Sim} in Figure~\ref{fig:llm_sim_prompt}.

\begin{figure}[h]
\centering

\begin{tcolorbox}[
    enhanced,
    breakable,
    width=\linewidth,
    colback=oldpaperyellow,
    colframe=black,
    arc=4mm,
    boxrule=0.4pt,
    left=6pt, right=6pt, top=6pt, bottom=6pt,
]

{\small\itshape

You are simulating \{domain\_expert\_desc\}, who is interacting with a data discovery system to explore insights from an enterprise dataset.\\

\{Depending on system:

- \textsc{Pneuma-Seeker}:  
  The system represents your information need as a set of target schemas and SQL statements that, if executed, will provide the answer. It can combine, transform, and reason over data to assist your exploration.\\
- FTS or \textsc{Pneuma-Retriever}:  
  The system only returns relevant tables based on your description. It does not infer your deeper intent, combine, or analyze data.
\}\\

Scenario:\\
- The system already has access to internal datasets.\\
- You (the simulated user) are familiar with the domain and have seen similar datasets before.\\
- You are not uploading new datasets or asking if they exist — you assume they do.\\

Possible eventual goal (unknown at start):\\
\{question\}\\

Behavior:\\
- Explore and refine your question step-by-step depending on the system's responses.\\
- Be vague or explore tangents, just as a curious analyst would.\\
- Only arrive at the specific question above if the system's output correctly leads you there.\\

Continue your role as the domain expert. This is the conversation so far (respond as if prompting the system directly):\\
YOU: \{initial\_broad\_prompt\}
}

\end{tcolorbox}

\caption{Prompt for \textsc{LLM\_Sim}}
\label{fig:llm_sim_prompt}
\end{figure}

\subsection{RQ1: Convergence}

Convergence means the \textsc{LLM Sim}'s active information need matches its latent information need. We introduce two metrics: \textbf{(1) percentage of convergence}, which is the proportion of benchmark questions for which \textsc{LLM Sim} converges, and \textbf{(2) median turns to convergence}, which is the median of the number of times \textsc{LLM Sim} has to prompt a system to achieve convergence (with an imposed limit of 15).

We compare \textsc{Pneuma-Seeker} with three baselines: BM25-based full-text search (FTS), \textsc{Pneuma-Retriever}, and \textsc{LlamaIndex}\footnote{\url{https://www.llamaindex.ai/}} (a representative RAG system). FTS and \textsc{Pneuma-Retriever} are static systems that only return tables, represented by their columns and sample rows. \textsc{LlamaIndex} adds an LLM on top of a top-$k$ vector retriever to interpret the retrieved data for \textsc{LLM Sim}.

\begin{figure}[h]
    \centering
    \includegraphics[width=1.0\linewidth]{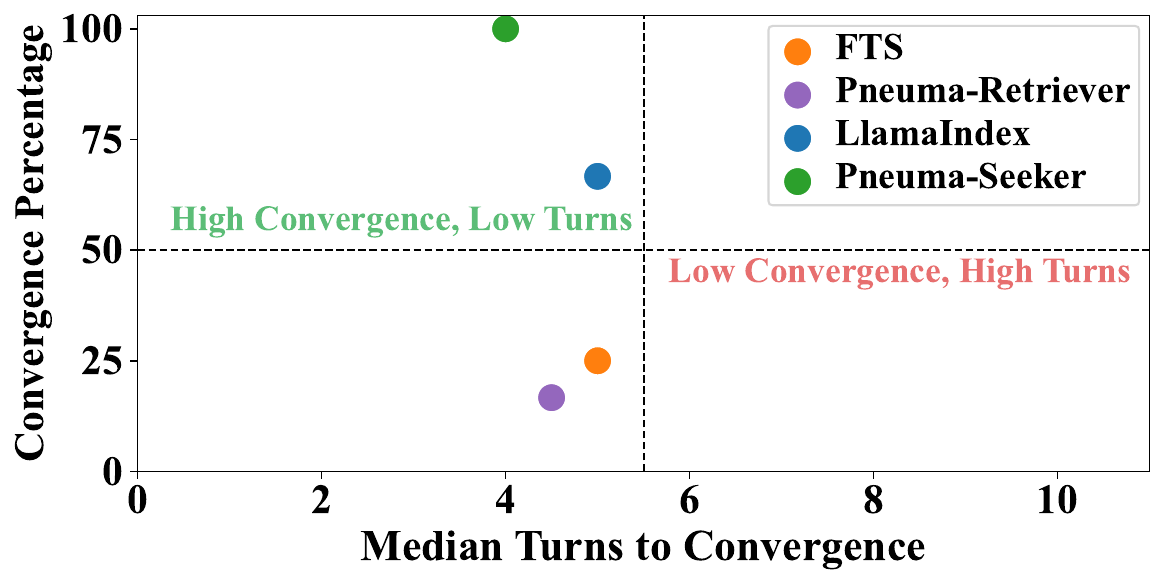}
    \caption{Comparison of Median Turns to Convergence vs. Convergence Percentage (Archeology Dataset)}
    \label{fig:archeo_rq1}
\end{figure}

\begin{figure}[h]
    \centering
    \includegraphics[width=1.0\linewidth]{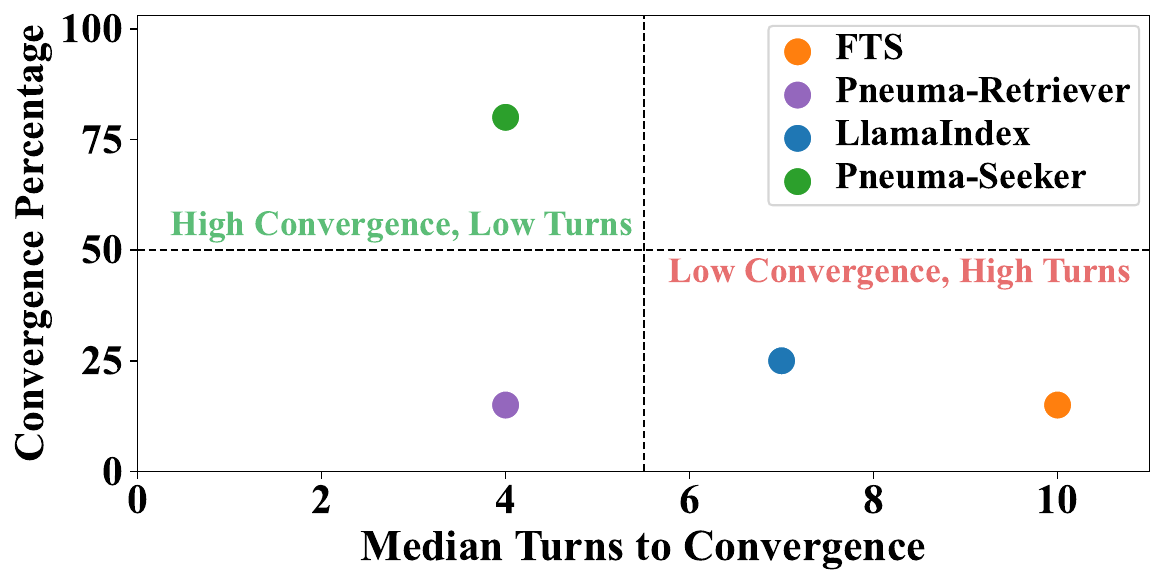}
    \caption{Comparison of Median Turns to Convergence vs. Convergence Percentage (Environment Dataset)}
    \label{fig:env_rq1}
\end{figure}

The results, as shown in Figure~\ref{fig:archeo_rq1} and Figure~\ref{fig:env_rq1}, indicate that \textsc{Pneuma-Seeker} consistently achieves the highest percentage of convergence and similar median turns to convergence as \textsc{LlamaIndex}. Both \textsc{Pneuma-Seeker} and \textsc{LlamaIndex} not only surface relevant data but also interpret it, contextualizing their responses to \textsc{LLM Sim}. Even though the initial queries from \textsc{LLM Sim} are more general, the systems surface relevant data and suggestions and communicate them back to \textsc{LLM Sim}. Subsequently, \textsc{LLM Sim} responds to the systems to explore further and move closer to uncovering its latent information need, and the systems adjust accordingly (e.g., by modifying the state).

On the other hand, both FTS and \textsc{Pneuma-Retriever} struggle to converge because \textsc{LLM Sim} has to interpret the results themselves. Additionally, even though \textsc{Pneuma-Retriever} can find the correct tables in almost all questions, \textsc{LLM Sim} can only observe sample rows to prevent hitting the context limit. Even with \verb|GPT-4o|'s 128k context limit, 2-3 turns are enough to exceed the limit with our datasets. In most of the questions, \textsc{LLM Sim} keeps trying to adjust its queries to get more specific information from the retrieved tables.

There is a latency trade-off between \textsc{Pneuma-Seeker} and the other systems. On average, \textsc{Pneuma-Seeker} takes 70.26 seconds to respond to a prompt, while FTS and \textsc{Pneuma-Retriever} answer almost instantaneously. In addition, \textsc{Pneuma-Seeker}'s LLM, \verb|OpenAI's O4-mini|, incurs $\$1.1$ and $\$4.4$ for every 1 million input and output tokens, respectively. For reference, we include the estimated average costs of interactions between \textsc{LLM Sim} and \textsc{Pneuma-Seeker} across several models in Table~\ref{tab:cost}.

\begin{table*}[t]
\centering
\caption{Estimated Average Token Usage and Costs Across Different LLMs}
\label{tab:cost}
\begin{tabular}{lcccccccccccccccccc}
\toprule
\textbf{Dataset} &
\textbf{Avg. In} &
\textbf{Avg. Out} &
\multicolumn{2}{c}{\texttt{Haiku 4.5}} &
\multicolumn{2}{c}{\textbf{\texttt{O4-mini}}} &
\multicolumn{2}{c}{\texttt{O3}} &
\multicolumn{2}{c}{\texttt{gpt-5.1}} &
\multicolumn{2}{c}{\texttt{Sonnet 4.5}} &
\multicolumn{2}{c}{\texttt{Opus 4.5}} \\
\cmidrule(lr){4-5}
\cmidrule(lr){6-7}
\cmidrule(lr){8-9}
\cmidrule(lr){10-11}
\cmidrule(lr){12-13}
\cmidrule(lr){14-15}
 & \textbf{Tokens} & \textbf{Tokens} &
In & Out &
\textbf{In} & \textbf{Out} &
In & Out &
In & Out &
In & Out &
In & Out \\
\midrule
Archeology  & 248{,}351 & 2{,}854
& \$0.25 & \$0.01
& \textbf{\$0.27} & \textbf{\$0.01} 
& \$0.50 & \$0.02
& \$0.31 & \$0.03
& \$1.49 & \$0.04
& \$1.24 & \$0.07 \\

Environment & 149{,}011 & 1{,}712
& \$0.15 & \$0.01
& \textbf{\$0.16} & \textbf{\$0.01}
& \$0.30 & \$0.01
& \$0.19 & \$0.02
& \$0.45 & \$0.03
& \$0.75 & \$0.04 \\
\bottomrule
\end{tabular}
\end{table*}

\subsection{RQ2: Accuracy}

Converging to the latent information need is not enough; users ultimately want to get accurate answers for their information needs. FTS and \textsc{Pneuma-Retriever} are not designed to provide answers, so we exclude them from RQ2 and include new baselines: \textsc{DS-Guru} and OpenAI's O3. \textsc{DS-Guru} is the Kramabench's reference framework, in which it instructs an LLM to decompose a question into a sequence of subtasks, reason through each step, and synthesize Python code implement the plan. We select the O3-based \textsc{DS-Guru}, as it is the best-performing one.

O3 is one of the best reasoning models right now with 200k context limit. For each benchmark question, we provide it with the whole relevant tables, so it has every necessary information to answer the question. However, we encountered context length exceeded errors with O3 in 6 out of 12 archaeology questions and 17 out of 20 environment questions. O3 answers none of the six archaeology questions correctly, but answers two environment questions correctly. Overall, passing all relevant context is still not a scalable approach.

For the remaining systems, the results are shown in Table~\ref{tab:convergence_rates}. \textsc{Pneuma-Seeker} outperforms all systems across all datasets. \textsc{Pneuma-Seeker} outperforms \textsc{DS-Guru}, even though \textsc{Pneuma-Seeker} uses a smaller and more cost-efficient model. \textsc{LlamaIndex}, on the other hand, does not answer any questions correctly because the questions require actual computation (e.g., computing average of a certain column), not just interpretation of the top-$k$ context.

\begin{table}[h]
\centering
\caption{Comparison of Accuracy across Datasets}
\label{tab:convergence_rates}
\begin{tabular}{lccc}
\toprule
\textbf{System} & \textbf{Archeology} & \textbf{Environment} \\
\midrule
\textsc{LlamaIndex} & 0.00\% & 0.00\% \\
\textsc{DS-Guru} (O3) & 25.00\% & 19.60\% \\
\textsc{Pneuma-Seeker} & \textbf{41.67\%} & \textbf{55.00\%} \\
\bottomrule
\end{tabular}
\end{table}

\section{Discussion}
In this section, we discuss lessons learned from building \textsc{Pneuma-Seeker} and explain our vision for the Pneuma project.

\subsection{Lessons Learned.}

We built \textsc{Pneuma-Seeker} over the course of the past year. We learned the following lessons that collectively motivated the architecture presented in this paper and directly informed the design of each component and their interactions.

\mypar{Dynamic pipeline enables better adaptability} Our early prototypes followed a fixed processing sequence. We can certainly adapt this pipeline to include more actions/tools and hence cover more use cases. However, we realized that evolving such pipelines requires an increasing amount of effort. As soon as we encounter a use case that is not covered by the system, e.g., requiring new actions or interactivity, we are forced to do an extensive re-engineering.

This happened repeatedly. For example, when we had to determine which actions are required or optional, whether to skip or restart some actions, and when we had to integrate new actions without destabilizing the existing pipeline. A static approach \emph{can} be adapted to handle these scenarios, but the engineering burden grows quickly as the system incorporates more and more actions and functionality.

Dynamic pipelines are not a silver bullet either: if a necessary capability is missing (e.g., semantic joins), the system will still fail. However, the key difference is that missing capabilities in a dynamic framework are \emph{localized}: once implemented, they naturally slot into the system's action space. In contrast, adding the same capability to a static pipeline necessitates revisiting the entire pipeline.

This realization aligns with the broader spirit of AI-enabled systems: instead of hard-coding rigid sequences, the system should select actions dynamically based on evolving state, available tools, and accumulated domain knowledge. It also resonates with the declarative philosophy of data processing: specifying \emph{what} is needed, not \emph{how} to accomplish it.

Existing systems such as \textsc{ReAcTable}~\cite{ReactTable2024} and \textsc{Chain-of-Table}~\cite{wang2024chainoftable} also adopt dynamic orchestration in a more specific setting: assuming a conventional tabular QA setting where a single, specific table is known upfront. In our setting, the system may need multiple tables and non-tabular data to satisfy users' information needs. Overall, factors such as handling user feedback, maintaining iterative alignment of the state, and integrating heterogeneous data sources all push strongly against static pipelines. These observations collectively led us away from predefined pipelines and toward \textsc{Conductor}’s flexible, state-driven dynamic orchestration.

\mypar{Context specialization is essential}
A key challenge in dynamic systems is the design of components themselves. We observed the importance of specializing context across different components (e.g., \textsc{Conductor} and \textsc{Materializer}). Beyond reducing the chance of hallucinations (since each component focuses narrowly on its role), specialization helps prevent exceeding context-window limits, which are more easily reached when the LLM is prompted with multiple different roles at once.

Looking ahead, although there is active research on extending context windows and future models likely will support larger windows, there is no guarantee that the model will effectively attend to (or meaningfully use) all its input tokens, even if they fit. Current models already have much longer context windows compared to those available just 3 years ago, but recent work still shows that LLMs' performance on tasks such as question answering degrades as context length grows, even when the relevant information is fully retrievable~\cite{context-length-hurts-llm}. Theoretical analyses such as~\cite{chen2025criticalattentionscalinglongcontext} also show that longer sequence length dilutes the model's ability to focus on specific tokens. Therefore, we argue that context specialization remains beneficial, especially as the system grows, e.g., with new actions and tools.

\mypar{Schema-first representation guides alignment}
Reifying information needs as a relational model provides a shared anchor that both the user and the system can collaboratively reason about. It gives users a concrete object to sanity-check the results, rather than relying purely on natural-language, user-facing messages. The state $(T, Q)$ becomes a structure that can be iteratively refined, corrected, or extended, reducing miscommunication.

Surfacing intermediate $(T, Q)$ states allows both users (and \textsc{LLM Sim}) to detect subtle misinterpretations early, before they propagate into later steps. Natural language alone is insufficient; users need visibility into the evolving data model to provide concrete, meaningful feedback. It also forces users to think concretely about what exactly they need. Designing effective ways to communicate this evolving state (and studying how users interpret and act on it) remains an important direction for future user studies.

\subsection{Our Vision}
We began the Pneuma project a year and a half ago, building on over five years of research in data discovery. Our first milestone was \textsc{Pneuma-Retriever}~\cite{BalakaPneuma2025}. In this paper, we have presented the current state of the project, which has since grown to address a critical bottleneck in data-centric organizations: as LLMs become increasingly capable and embedded in data workflows, the main challenge shifts to helping users articulate what they want to get out of data. Our vision is a system that treats this articulation process as a first-class concern. This is reflected in our design through the reification of information needs as relational schemas tailored to the user's active intent and constructed on-the-fly.

We highlight an important emergent effect of \textsc{Pneuma-Seeker}'s design. By prompting users to clearly express their goals, disclose assumptions, and externalize their mental models---a behavior increasingly familiar in LLM-driven interfaces---\textsc{Pneuma-Seeker} captures what is often tacit knowledge. Within organizations, this ``tribal knowledge'' represents a collective brain trust. If made searchable and persistent, it could fulfill the vision of internal data markets we proposed in earlier work~\cite{datamarkets}. Such markets would enable organizations to extract far greater value from their data assets, going well beyond the immediate gains in discovery and preparation addressed in this paper. Over the past several years, we have made progress toward that vision, but extracting tribal knowledge remains a hard-to-crack barrier. The Pneuma project is our latest and most focused attempt to lower that barrier and in doing so, to unlock latent value from organizational data.

% \subsection{Future Work}

% \noteltf{I will try to list things that I want to implement (all of these are also available in Section 3)}
% \begin{enumerate}
%     \item Allow user to supply their data outside of their organizational data indexed in Pneuma-Seeker ('Bring Your Own Data'), similar to how in ChatGPT, we can upload documents.
%     \item Add fine-grained provenance tracking, essentially for each value in the resulting documents ($Q$), where does it come from, what transformation was done on the underlying data to produce this value.
%     \item Automatically capture knowledge from user interaction (e.g., the fact that there are direct and indirect tariff, or tariff is relative)
%     \item Explore interaction models in which Conductor provides incremental feedback to Materializer after every n operations.
% \end{enumerate}

% \printbibliography
\bibliographystyle{plainnat}   % or abbrvnat, unsrtnat
\bibliography{main}

\end{document}